\documentclass{proc}

\usepackage{amssymb}
\usepackage{stfloats}
\usepackage[dvips]{color}
\usepackage{graphicx}
\usepackage{amsmath}
\usepackage{multirow,subfigure}
\usepackage{bm}
\usepackage{booktabs}
\usepackage{times}
\usepackage[english]{babel}
\usepackage{cuted}

\makeatletter
\def\@maketitle{%
  \vbox to 2.3in{%
    \hsize\textwidth
    \linewidth\hsize
    \vspace*{1.5cm}
    \centering
    {\bfseries\LARGE \@title \par}
    \vskip 2em
    {\large \begin{tabular}[t]{c}\@author \end{tabular}\par}
    \vfill}    \vspace*{1.0cm}
}
\renewcommand\section{\@startsection {section}{1}{\z@}%
     {.7\baselineskip plus\baselineskip}{.5\baselineskip}
                                   {\normalfont\Large\bfseries}}
\renewcommand\section{\@startsection {section}{1}{\z@}%
      {.5\baselineskip\@plus.7\baselineskip}{.3\baselineskip}%
                                   {\normalfont\Large\bfseries}}
\renewcommand\subsection{\@startsection{subsection}{2}{\z@}%
       {.5\baselineskip\@plus.7\baselineskip}{.3\baselineskip}%
                                   {\normalfont\large\bfseries}}
\renewcommand\subsubsection{\@startsection{subsubsection}{3}{\z@}%
      {.5\baselineskip\@plus.7\baselineskip}{.3\baselineskip}%
                                     {\normalfont\normalsize\bfseries}}
\makeatother

\renewenvironment{abstract}%
  {\normalfont
    \list{}{\labelwidth0pt
      \leftmargin0pt \rightmargin\leftmargin
      \listparindent\parindent \itemindent0pt
      \parsep0pt
      
    }%
    \item[\hskip\labelsep\bfseries\abstractname\enspace --] \itshape%
}{%
  \endlist}

\newcommand{\keywordsname}{Keywords}
\newenvironment{keywords}%
  {\normalfont
    \list{}{\labelwidth0pt
      \leftmargin0pt \rightmargin\leftmargin
      \listparindent\parindent \itemindent0pt
      \parsep0pt
      }%
    \item[\hskip\labelsep\bfseries\keywordsname:]}{\endlist}
  
\begin{document}

\title{Approximate Propagation of both Epistemic and Aleatory Uncertainty through Dynamic Systems}



\author{\begin{tabular}{ccccccc}
\bf Gabriel Terejanu$^{a}$ && \bf Puneet Singla$^{b}$ & &\bf Tarunraj Singh$^b$ & & \bf Peter D. Scott$^a$\\
terejanu@buffalo.edu &&psingla@buffalo.edu && tsingh@buffalo.edu && peter@buffalo.edu
\end{tabular}\\
$^a$\textsl{Department of Computer Science \& Engineering}\\
$^b$\textsl{Department of Mechanical \& Aerospace Engineering}\\
\textsl{University at Buffalo, Buffalo, NY-14260.}} \maketitle

\selectlanguage{english}

\begin{abstract}
When ignorance due to the lack of knowledge, modeled as epistemic uncertainty using Dempster-Shafer structures on closed intervals, is present in the model parameters, a new uncertainty propagation method is necessary to propagate both aleatory and epistemic uncertainty. The new framework proposed here, combines both epistemic and aleatory uncertainty into a second-order uncertainty representation which is propagated through a dynamic system driven by white noise. First, a finite parametrization is chosen to model the aleatory uncertainty by choosing a representative approximation to the probability density function conditioned on epistemic variables. The epistemic uncertainty is then propagated through the moment evolution equations of the conditional probability density function. This way we are able to model the ignorance when the knowledge about the system is incomplete. The output of the system is a Dempster-Shafer structure on sets of cumulative distributions which can be combined using different rules of combination and eventually transformed into a singleton cumulative distribution function using Smets' pignistic transformation when decision making is needed.
\end{abstract}


\section{Introduction}
\label{sec:Introduction}
The paper presents a novel framework to characterize the response of stochastic dynamic systems with model parameters governed by epistemic uncertainty. We distinguish between \textit{aleatory uncertainty} which arises due to the stochastic behavior of the system and \textit{epistemic uncertainty} which is used to model the ignorance in the model parameters. In this last case we do not know if the parameter in discussion is random or not, and if it is random what is its underlying probability distribution~\cite{Oberkampf2004}. 

Ferson and Ginzburg \cite{Ferson1996} argues that when dealing with propagating uncertainty through mathematical models, different calculation methods are required to propagate aleatory and epistemic uncertainty. However, the segregation of both types of uncertainty during propagation, as well as the accurate propagation of either one of them are two challenges in creating a method to propagate both epistemic and aleatory uncertainty. 

Finding only the response of a stochastic dynamic system with no uncertainty in the parameters, is a very active area of research. The problem in this case is to solve for the evolution of the probability density function (pdf), $p(t,x)$, corresponding to the state $x$ of the dynamic system. The evolution of the pdf is governed by the Fokker-Planck-Kolmogorov (FPK) equation with analytical solution known only for stationary pdfs of a limited class of dynamic systems \cite{Fuller1969,Risken1989}. Different approximate methods exists in the literature to represent the pdf with a finite number of parameters such as finite-difference techniques, Monte Carlo methods, Gaussian Mixtures models \cite{Terejanu2008} and Gaussian closure methods.

These methods provide approximations to the pdf when the prior pdf is precisely known, the process noise is perfectly characterized and the parameters in the model are exactly known. This is also known as the \textit{Bayesian dogma of precision} \cite{Walley1991}, however, in practice, these precise values are difficult to obtain due to the amount of information available, incomplete knowledge of the system or systematic underestimation of uncertainty which arises in the elicitation process. This becomes of great importance when doing hazard risk assessment and decision making. 


The framework proposed here, propagates both epistemic and aleatory uncertainty through a dynamic system driven by white noise using a two-level hierarchical model. On the first level we model the aleatory uncertainty by choosing a representative approximation for the pdf with a finite number of parameters. In this paper we model the pdf conditioned on epistemic variables as a Gaussian distribution. 


On the second level is the epistemic uncertainty that we have in the moments of the stochastic solution, in this case the first two moments. The epistemic uncertainty is modeled in this work using finite Dempster-Shafer structures where the focal elements are closed intervals. The closed intervals are propagated through the moment evolution equations using a recent proposed method based on Polynomial Chaos - Bernstein Form \cite{Terejanu2010}. Thus the framework provides the means to propagate both epistemic and aleatory uncertainty separately \cite{Ferson1996} as well as combine them when a decision has to be made. 

After the propagation we obtain finite Dempster-Shafer structures, only this time the focal elements are sets of distributions represented by probability boxes \cite{Ferson2003}. The structures obtained characterize the uncertainty in the quantity of interest, namely the cumulative distribution function (cdf) of the response of the system, and they be used with classic decision theory by using the pignistic transformation \cite{Smets1994} whenever a decision is required. 

In Section \ref{sec:Problem} the problem to be solved is defined, followed by a background of Demspter-Shafer structures on closed intervals and decision making under ignorance in Section \ref{sec:Background}. The approximate uncertainty propagation framework is developed in Section \ref{Proposal} with resolutions for the two posed problems. In Section \ref{sec:Results} a proof of concept example is presented and the conclusions and future work are discussed in Section~\ref{sec:Conclusions}.

\section{Problem Statement}
\label{sec:Problem}
Consider the following first order model with stochastic forcing and uncertain initial condition modeled using a Gaussian pdf:
\begin{eqnarray} \label{probModel}
\dot{x}(t) + f(x,\bm\alpha,t) &=& \Gamma(t) \\
p(t_0, x_0) &=& \mathcal{N}(x_0;\mu_0,\sigma_0^2)
\end{eqnarray}

The Gaussian white noise process $\Gamma(t)$ has the autocorrelation function $E[\Gamma(t)\Gamma(\tau)] = q^2 \delta(t-\tau)$, and $\bm\alpha$ is a vector of $n$ parameters.

The variables of the model can be segregated in two sets: the set of \textit{aleatory variables} given by $\mathbf{a}(t) = \{x(t), \Gamma(t)\}$ and the set of independent \textit{epistemic variables} given by $\mathbf{e}(t) = [ \mathbf{e}_S, \mathbf{e}_D(t) ]$, where $\mathbf{e}_D(t) = \{ m_1(t), m_2(t) \ldots m_k(t) \}$ is a time-variant vector of epistemic moments used to characterize the probability distribution of the response, $p(t,x|\mathbf{e}_D)$, and $\mathbf{e}_S = [\bm\alpha, q^2 ]$ is a time-invariant vector of epistemic variables. Given the initial condition, the first two moments are given by $m_1(0) = \mu_0$ and $m_2(0) = \sigma_0^2 + \mu_0^2$. 

If a part of the model parameters, $\bm\alpha$, are characterized by aleatory uncertainty then they can be treated as additional state variables and they can be augmented to the aleatory vector $\mathbf{a}(t)$. The evidence about the model parameters in $\mathbf{e}(t)$ is modeled here using Dempster-Shafer structures on closed intervals and the uncertainty of the aleatory variables in $\mathbf{a}(t)$ is quantified by probability distributions. 

When the model parameters are known, the response of the system is characterized only by aleatory uncertainty which is represented here by the cumulative distribution function,
\begin{eqnarray}
P(X \le x_f | \mathbf{e}_0) = \int_{-\infty}^{x_f} p(t, x | \mathbf{e}_0) \mathrm{d}x
\end{eqnarray}
where $\mathbf{e}_0 = \mathbf{e}(0) = [\mathbf{e}_S, \mathbf{e}_D(0)]$ and $p(t, x | \mathbf{e}_0)$ is obtained in general by solving the FPK equation. 

Given the uncertain model parameters, we are interested in solving the following three problems:
\begin{enumerate}
	\item The main objective is to find the induced Dempster-Shafer structure about the response of the system, $m[P(X \le x_f | \mathbf{e}_0)]$. Hence, we are looking to determine the focal elements of the response as well as their corresponding probability masses.
	\item Construct a cumulative density function using the pignistic transformation in order to use the expected utility theory for decision making.
\end{enumerate}

The last problem represents a secondary objective and is included here in order to present the practicality of the approach.

\section{Background}
\label{sec:Background}
A DS-structure on closed intervals is a collection of interval-valued focal elements and their associated basic probability assignments \cite{Ferson2003}. We write the uncertainty of the epistemic variable $x$ is represented by the following DS-structure as:
\begin{eqnarray}\label{DSstruct}
x \sim \big\{~([\underline{x}_1,\overline{x}_1],p_1) ~,~ ([\underline{x}_2,\overline{x}_2],p_2) ~,~ \ldots ~([\underline{x}_n,\overline{x}_n],p_n) ~\big\}
\end{eqnarray}

Given two nondecreasing functions $\overline{F}$ and $\underline{F}$, where $\overline{F}, \underline{F} : \mathbb{R} \rightarrow [0,1]$ and $\underline{F}(x) \le \overline{F}(x)$ for all $x \in \mathbb{R}$, we can represent the imprecision in the cumulative distribution function, $F(x)=P(X\le x)$, by the probability box (p-box) $[\underline{F},\overline{F}]$ as follows: $\underline{F}(x) \le F(x) \le \overline{F}(x)$ \cite{Ferson2003}.

A Dempster-Shafer structure on closed intervals can induce a unique p-box, while the inverse is not uniquely determined. Many Dempster-Shafer structures exist for the same p-box. Given the following body of evidence, Eq.\eqref{DSstruct}, the cumulative belief function (CBF) and the cumulative plausibility function (CPF) are defined by:
\begin{eqnarray}\label{cdf}
CBF(x) = \underline{F}(x) = \sum_{\overline{x}_i \le x} p_i \\
CPF(x) = \overline{F}(x) = \sum_{\underline{x}_i \le x} p_i 
\end{eqnarray} 

Thus the cumulative distribution function is bounded as follows:
\begin{eqnarray}
CBF(x) \le P(X \le x) \le CPF(x)
\end{eqnarray} 

In order to compute expectations using the interval-based belief functions, one needs to build a probability density function given the belief structure. Based on Smets' pignistic transformation one can define the pignistic probability density function $p_{Bet}(x)$ \cite{PetitRenaud2004} as a finite mixture of continuous uniform distributions:
\begin{eqnarray}\label{pignistic_pdf}
p_{Bet}(x) = \sum_{i = 1}^{n} \frac{p_i}{\overline{x}_i - \underline{x}_i} I(x,[\underline{x}_i, \overline{x}_i])
\end{eqnarray}
where the indicator function $I(x,[\underline{x}_i, \overline{x}_i])$ is given by:
\[
I(x,[\underline{x}_i, \overline{x}_i]) = \left \{
\begin{array}{cc}
    1, & \mathrm{for}~x \in [\underline{x}_i, \overline{x}_i] \\
    0, & \mathrm{otherwise} \\
\end{array}
\right.
\]

Using the above defined pignistic probability density function $p_{Bet}(x)$ one can compute the expectations needed in the decision making process. Example the expected value of $x$ given the pignistic probability density function si given by:
\begin{eqnarray}
\mathrm{E}_{Bet}(X) = \int_{-\infty}^{+\infty} x p_{Bet}(x) \mathrm{d}x = \frac{1}{2} \sum_{i=1}^n p_i  (\underline{x}_i+ \overline{x}_i)
\end{eqnarray}

The pignistic transformation constructs a singleton pdf and makes the expected utility theory applicable, however it ignores the ignorance \cite{Haenni2003}. One possible way to incorporate the ignorance into the decision process is to construct a scalar measures which quantifies the total amount of ignorance and take a decision only if the level of ignorance is relatively low. Define the normalized integral of the degree of ignorance:
\begin{eqnarray}\label{NIDI}
NIDI = \frac{1}{\overline{x}-\underline{x}} \int_{\underline{x}}^{\overline{x}} [CPF(x)-CBF(x)] \mathrm{d}x 
\end{eqnarray}
where $\underline{x} = \min_{1 \le i \le n} \underline{x}_i$ and $\overline{x} = \max_{1 \le i \le n} \overline{x}_i$.
The $NIDI$ is a scalar measure with values ranging between $0.0$ and $1.0$ and it summarizes the confidence in the pignistic pdf. A value equal with $1.0$ denotes that we are dealing with interval uncertainty where we only know the bounds of the variable, and a value closed to $0.0$ means that we know the pdf of $x$ precisely. Thus, one can be more confident in applying the expected utility theory if the amount of ignorance if low (eq. less than $0.1$) and can try to defer the decision making and gather more evidence if the level of ignorance is not tolerable.

%

\section{Proposed Approach}
\label{Proposal}

\subsection{Time evolution of moments}
\label{sec:Moments}
In order to keep both types of uncertainty segregated during propagation, one can use the moment evolution equations associated with Eq.\eqref{probModel}, which is a set of ordinary differential equations (ODE) governed only by epistemic uncertainty. Thus, after the applying the It$\hat{\mathrm{o}}$'s lemma to Eq.\eqref{probModel} \cite{Jazwinski1970}, the time evolution of the moments conditioned on the epistemic variables is given by
\small
\begin{eqnarray} \label{jaz_mom}
\frac{\mathrm{dE}[\varphi(x)|\mathbf{e}_0]}{\mathrm{d}t} = -\mathrm{E}[\varphi_x(x)f(x,\bm\alpha,t)|\mathbf{e}_0] + \frac{1}{2}\mathrm{E}[q^2\varphi_{xx}(x)|\mathbf{e}_0]
\end{eqnarray}
\normalsize
where $\varphi(x)=x^k$. 

The nonlinearity of the model is assumed to be polynomial, however the general case can be handled using numerical integration schemes such as Gaussian quadrature. Thus, when the nonlinearity in the model is polynomial,
\begin{eqnarray}
f(x,\bm\alpha,t) &=& \sum_{i=1}^n \alpha_i x^i~,
\end{eqnarray}
the final form for the evolution of the moments conditioned on the epistemic variables, is given by
\small
\begin{eqnarray}\label{final_mom}
\dot{m}_k|\mathbf{e}_0 = -k \sum_{i=1}^n \alpha_i m_{i+k-1}|\mathbf{e}_0 + \frac{1}{2}k(k-1)q^2m_{k-2}|\mathbf{e}_0~.
\end{eqnarray}
\normalsize

Observe, that in Eq.\eqref{final_mom} the lower order moments depend on the higher order moments, thus we have an infinite hierarchy of moment equations. To truncate the infinite chain, one can use moment closure schemes which assume a specific class of probability distributions to express higher order moments in terms of lower order ones. 

In this paper we are using the Gaussian closure method, hence we are interested in propagating only the first two moments, $m_1$ and $m_2$, under the assumption that $p(t,x|\mathbf{e}_D) \approx \mathcal{N}(x ; \mu, \sigma^2|\mathbf{e}_D)$. To express the moments of order $k>2$ in terms of the lower order moments one can use the following relation \cite{Abramowitz1972}:
\small
\begin{eqnarray}
m_k &=& \mathrm{E}\left( (X-\mu)^k \right) - \sum_{i=0}^{k-1} (-1)^{k-i} \binom{k}{i} m_i \mu^{k-i}
\end{eqnarray}
\normalsize
where $m_0 = 1$, $\mu = m_1$, $\sigma^2 = m_2 - m_1^2$, and the central moments of the Gaussian distribution are given by
\[
\mathrm{E}\left( (X-\mu)^k \right) = \left \{
\begin{array}{cc}
    \sigma^k \frac{k!}{2^{k/2} (k/2)!}, & k~\mathrm{is~even} \\
    0, & k~\mathrm{is~odd} \\
\end{array}
\right.
\]

After using the Gaussian closure, the time-variant vector of epistemic variables is given only by the first two moments, $\mathbf{e}_D(t) = [m_1(t), m_2(t)]^T$, and the system of ODEs \eqref{final_mom} is transformed into the following system:
\small
\begin{eqnarray}\label{final_form}
\dot{\mathbf{e}}_D &=& \mathbf{g}(\mathbf{e}_D, \mathbf{e}_S) \\
&& \mathbf{e}_D(0) \sim \big\{~([\underline{\mathbf{e}}_{D1}(0),\overline{\mathbf{e}}_{D1}(0)],p_{D1}) ~,~ \nonumber\\
&& \hspace{0.6in} ([\underline{\mathbf{e}}_{D2}(0),\overline{\mathbf{e}}_{D2}(0)],p_{D2}) ~,~ \ldots ~\big\} \nonumber \\
&& \mathbf{e}_S \sim \big\{~([\underline{\mathbf{e}}_{S1},\overline{\mathbf{e}}_{S1}],p_{S1}) ~,~ ([\underline{\mathbf{e}}_{S2},\overline{\mathbf{e}}_{S2}],p_{S2}) ~,~ \ldots ~\big\} \nonumber
\end{eqnarray}
\normalsize

Under the Gaussian closure assumption, $p(t,x|\mathbf{e}_0) = p(t,x|\mathbf{e}_D)$ where $\mathbf{e}_D$ is obtain given $\mathbf{e}_0$ and the Eq. \eqref{final_form}.

We are interested in finding the induced Dempster-Shafer structure in $\mathbf{e}_D(t)$ at time $t > 0$. Given the knowledge about elements of $\mathbf{e}_0$ as independent DS structures on closed intervals, we need to find the belief structure associated with $\mathbf{e}_D(t)$. The DS-structure describing the uncertainty for $\mathbf{e}_D(t)$ can be obtained using Yager's convolution rule for DS-structures under the assumption of independence \cite{Yager1986}:
\begin{equation}\label{mapping}
m[\mathbf{e}_D(t)](\mathbf{C}_{ij}) = \mathop{ \mathop { \sum_{ \dot{\mathbf{e}}_D = \mathbf{g}(\mathbf{e}_D, \mathbf{e}_S)} }_{ \mathbf{e}_D(0) \in \mathbf{A}_i } }_{ \mathbf{e}_S \in \mathbf{B}_j } \underbrace{m[\mathbf{e}_D(0)](\mathbf{A}_i)}_{p_{Di}} \underbrace{m[\mathbf{e}_S](\mathbf{B}_j)}_{p_{Sj}}
\end{equation}
where $\mathbf{C}_{ij}=[\underline{\mathbf{e}}_{Dij}(t),\overline{\mathbf{e}}_{Dij}(t)], \mathbf{A}_i = [\underline{\mathbf{e}}_{Di}(0),\overline{\mathbf{e}}_{Di}(0)]$ and $\mathbf{B}_j = [\underline{\mathbf{e}}_{Sj},\overline{\mathbf{e}}_{Sj}]$. Here the only unknown is the collection of focal elements $\mathbf{C}_{ij}$ of $\mathbf{e}_D(t)$ .

Thus the problem of finding the mapping of a body of evidence on closed intervals is reduced to interval propagation \cite{Limbourg2008}. This problem can be solved using the advanced techniques developed in the interval analysis field \cite{Jaulin2001}. However, due to the dependence problem the obtained bounds are conservative which is detrimental to the belief structure, since the evidence is assigned automatically to other elements which are not in the body of evidence. This problem becomes more acute when the uncertainty has to be propagated over a period of time. 

The dependence problem in interval arithmetics can be avoided by using Taylor models and the remainder differential algebra for bounding the range of the response of ordinary differential equations under both initial value and parametric interval uncertainty \cite{Nedialkov2001, Berz1998, Lin2007}. In Ref.\cite{Enszer} are presented several examples in propagating uncertainties represented by probability boxes using Taylor model methods and interval arithmetics. Also an excellent review on interval methods for initial value problems is presented in Ref.\cite{Nedialkov1999}.

A method based on polynomial chaos and the Bernstein form is presented in Ref.\cite{Terejanu2010} to propagate DS-structures on closed intervals through nonlinear functions. Given a function of random variables with compact support probability distributions, the intuition is to quantify the uncertainty in the response using polynomial chaos expansion and discard all the information provided about the randomness of the output and extract only the bounds of its compact support. To solve for the bounding range of polynomials, we have proposed to transform the polynomial chaos expansion into the Bernstein form, and use the range enclosure property of Bernstein polynomials to find the minimum and maximum value of the response \cite{Cargo1966}.

The PCE is mathematically attractive due to the functional representations of the stochastic variables. It
separates the deterministic part in the polynomial coefficients and the stochastic part in the orthogonal polynomial basis. This becomes particularly useful in characterizing the uncertainty of the response of a dynamical system represented by ordinary differential equations with uncertain parameters such as in Eq.\eqref{final_form}. The result is a set of deterministic differential equations which can be solved numerically to obtain the evolution of the polynomial coefficients.

\subsection{Interval Uncertainty propagation through ODEs using PCE and the Bernstein Form}
\label{sec:PCEbernstein}
Given the initial value problem in Eq.\eqref{mapping} with interval-valued initial condition and parameters,
\begin{eqnarray} \label{PCEmodel}
\dot{e}_D^k &=& g_k(\mathbf{e}_D, \mathbf{e}_S) \\
&& \mathbf{e}_D(0) \in [\underline{\mathbf{A}},\overline{\mathbf{A}}] \nonumber \\
&& \mathbf{e}_S \in [\underline{\mathbf{B}},\overline{\mathbf{B}}] \nonumber
\end{eqnarray}
our goal is to approximate the range of the state variables $e_D^k$ at time $t$, where $e_D^k$ is the $k$-th component of $\mathbf{e}_D(t)$. Here, intervals of general type have been used in order to reserve the indexes for later use.

The problem can be transformed into finding the stochastic response under the assumption that both the initial condition and the model parameters are uniformly distributed. Thus we define $e_D^i(0) \sim \mathcal{U}(\underline{A}^i,\overline{A}^i)$ and $e_S^j \sim \mathcal{U}(\underline{B}^j,\overline{B}^j)$, where $e_D^i(0)$ and $e_S^j$ is the $i$-th component of $\mathbf{e}_D(0)$ and the $j$-th component of $\mathbf{e}_S$ respectively.

We expand both the uncertain arguments and the response of the system using the finite dimensional Wiener-Askey polynomial chaos \cite{Xiu2002}:
\begin{eqnarray}
e_S^l &=& \sum_{j=0}^{p-1} e_{Sj}^l \psi_j(\xi_l) \quad \mathrm{where} \quad \xi_l \sim \mathcal{U}(-1,1) \label{pol1} \\
e_D^k &=& \sum_{i=0}^{P-1} e_{Di}^k \bm\psi_i(\bm\xi) \quad \mathrm{where} \quad P = \frac{(r+p)!}{r!p!} \label{pol2}
\end{eqnarray}

Here $r$ is the number of uncertain input variables and is equal with the sum of the size of the vector $\mathbf{e}_D(0)$ and $\mathbf{e}_S$, and $p$ is the order of the polynomial expansion. The basis function $\psi_j$ is the $j$-th degree Legendre polynomial and $\bm\psi_i$ is a multidimensional Legendre polynomial and the polynomial coefficients are initialize such that they match their initial uniform assumption.

We are interested in finding the polynomial coefficients $e_{Di}^k$ of the response after $t$ sec. Substituting Eqs.\eqref{pol1}-\eqref{pol2} in Eq.\eqref{PCEmodel} and using the Galerkin projection and the orthogonality property of the polynomials one obtains a system of $P$ deterministic differential equations which can be solved numerically to obtain the PC expansion coefficients of the $k$-th moment.
\begin{eqnarray}\label{coeff}
\dot{e}_{Di}^k = \frac{< g_k, \bm\psi_k >}{<\bm\psi_k^2>}
\end{eqnarray}
where $<.~,~.>$ represents the inner product operator and can be evaluated in general using sampling or quadrature techniques. In this particular case since the nonlinearity is of polynomial type, the inner product between different Legendre basis functions can be computed \textit{a priori}, speeding this way the numerical integration of the ODE.

After $t$ sec, by integrating Eq.\eqref{coeff}, on obtains the polynomial coefficients which define the stochastic response of the system. However, we are only interested in finding the bounds that enclose the response of the system. In Ref.\cite{Terejanu2010} it is shown that by bringing the polynomial chaos expansion, Eq.\eqref{pol2}, to a Bernstein form using the Garloff's method \cite{Garloff1985}, one can efficiently find the range of the compact support thanks to the enclosing property of Bernstein polynomials. 
\subsection{DS structures on Probability Sets}
\label{sec:ProbSets}
By propagating the DS structures through the moment evolution equations we obtain, at time $t$, an induced DS structure for the moments that characterize the probability distribution of the response. Since in this paper we have chosen to approximate the conditional pdf of the response using Gaussian density functions, we are only interested in the DS structures for the first two central moments: the mean, $\mu$, and the variance, $\sigma^2$. Thus, using Eq.\eqref{mapping} the following DS structure is obtained for the first two moments:
\begin{eqnarray}\label{DSmom}
\mathbf{e}_D(t) &\sim& \big\{~([\underline{\mathbf{e}}_{D1}(t),\overline{\mathbf{e}}_{D1}(t)],p_{D1}) ~,~ \nonumber \\
&& \quad ([\underline{\mathbf{e}}_{D2}(t),\overline{\mathbf{e}}_{D2}(t)],p_{D2}) ~,~ \ldots ~\big\}
\end{eqnarray}

For each focal element we obtain a pair of two intervals that bound the range of the mean and the variance of a Gaussian density function.

Consider now the following Gaussian density function, $\mathcal{N}(x;\mu,\sigma^2)$, with uncertain parameters, $\mu$ and $\sigma^2$, given by two intervals: $[\underline{\mu},\overline{\mu}]$ and $[\underline{\sigma^2},\overline{\sigma^2}]$. Let us denote the cumulative distribution function $N(x_f;\mu,\sigma^2) = \int_{-\infty}^{x_f} \mathcal{N}(x;\mu,\sigma^2) \mathrm{d}x$. In Ref.\cite{Zhang2003} it is shown that all the normal cdfs are bounded by two functions, $\underline{N}(x_f) \le N(x_f;\mu,\sigma^2) \le \overline{N}(x_f)$, that can be computed analytically. 
\small
\begin{eqnarray}\label{envelope_prop}
\underline{N}(x_f) = \left\{
    \begin{array}{ll}
        N(x_f;\underline{\mu},\underline{\sigma^2}) \quad x_f \ge \underline{\mu} \\
        N(x_f;\underline{\mu},\overline{\sigma^2}) \quad x_f < \underline{\mu}
    \end{array}
\right. \\
\overline{N}(x_f) = \left\{
    \begin{array}{ll}
        N(x_f;\overline{\mu},\overline{\sigma^2}) \quad x_f \ge \overline{\mu} \\
        N(x_f;\overline{\mu},\underline{\sigma^2}) \quad x_f < \overline{\mu}
    \end{array}
\right. \nonumber
\end{eqnarray}
\normalsize

%

Using the above envelope property and the DS structure in Eq.\eqref{DSmom}, we obtain the following induced DS structure for the response of the system, 
\small
\begin{eqnarray}\label{finalDS}
P(X \le x_f | \mathbf{e}_0) &\sim& \bigg\{\big( [\underline{N}_1(x_f),\overline{N}_1(x_f)] ,p_{D1}\big) ~,~  \\
&& \quad ~\big( [\underline{N}_2(x_f),\overline{N}_2(x_f)] ,p_{D2} \big) ~,~ \ldots \bigg\} \nonumber
\end{eqnarray}
\normalsize

Thus, we are modeling a system with second order uncertainty and our credal set is defined as a DS structure over p-boxes. Having a structure like this makes the problem of decision making difficult, since we are dealing with many envelopes of cdfs instead of a singleton cdf or just an imprecise probability represented using a p-box. However, the following subsection presents how the pignistic transformation can be used whenever decision making is needed.

\subsection{Constructing a singleton CDF and Decision Making using DS structures on Probability Sets}
\label{sec:DMDSPS}
In order to compute expectations using the interval-based belief functions, one needs to build a probability density function given the belief structure. Based on Smets' pignistic transformation one can define the pignistic probability density function $p_{Bet}(x)$ \cite{PetitRenaud2004} as a finite mixture of continuous uniform distributions. Thus, given the DS structure in Eq.\eqref{finalDS}, we can construct a singleton cdf using the pignistic transformation as shown in Eq.\ref{pigCdf}. 
\begin{eqnarray}\label{pigCdf}
P_{Bet}(X \le x_f) &=& E_{Bet}\bigg( P(X \le x_f | \mathbf{e}_0) \bigg) \\
&=& \frac{1}{2} \sum_{i=1}^{n_D} \bigg(\underline{N}_i(x_f) + \overline{N}_i(x_f) \bigg) p_{Di} \nonumber
\end{eqnarray}

Here, $n_D$ is the total number of focal elements in the body of evidence Eq.\eqref{finalDS}. Notice that this is different from the law of total probability where a probability density function is constructed for the epistemic variables either $\mathbf{e}_0$ or $\mathbf{e}_D$. In both cases the pignistic transformation is applied much earlier in computing the quantity of interest. 

Furthermore one can construct an \textit{ignorance function} by applying Eq.\eqref{NIDI} for each $x_f$. This associates a point-wise measure of confidence in constructing the pignistic cdf.

\begin{strip}
\rule[0in]{3.3in}{.001in}
\small
\begin{eqnarray} \label{igorance_fun}
IgF(x_f) &=& \int_{ \underline{N}(x_f)}^{\overline{N}(x_f)} \bigg[ CPF_{P(X \le x_f | \mathbf{e}_0)}(z) - CBF_{P(X \le x_f | \mathbf{e}_0)}(z)\bigg] \mathrm{d}z \\
\mathrm{where}&&\underline{N}(x_f) = \min_{1\le i\le n_D} \underline{N}_i(x_f) \quad \mathrm{and} \quad \overline{N}(x_f) = \max_{1\le i\le n_D} \overline{N}_i(x_f) \nonumber \\
&&CPF_{P(X \le x_f | \mathbf{e}_0)}(z) =\sum_{\underline{N}_i(x_f) \le z} p_Di \nonumber \\
&&CBF_{P(X \le x_f | \mathbf{e}_0)}(z) =\sum_{\overline{N}_i(x_f) \le z} p_Di \nonumber
\end{eqnarray}
\normalsize
\begin{flushright}
\rule[0in]{3.3in}{.001in}
\end{flushright}
\end{strip}

Similar to Eq.\eqref{NIDI}, we can construct a scalar measure to summarize the total amount of ignorance by integrating the above ignorance function, 
\begin{eqnarray} \label{ig_meas}
NIigF = \frac{1}{x_{\max} - x_{\min}} \int_{x_{\min}}^{x_{\max}} IgF(x_f) \mathrm{d}x_f ~.
\end{eqnarray}

Given the pignistic cdf, $x_{\min}$ and $x_{\max}$ are the $p$th and $(100-p)$th percentiles respectively (eq. $p = 0.05$). The $NIigF$ is a number between $0.0$  and $1.0$ and can be used similarly as $NIDI$, in Section \ref{sec:Background}, to make decisions using the pignistic cdf in Eq. \eqref{pigCdf}, if it's value is small (eq. less than $0.1$) otherwise defer the decision if it is permitted and gather more evidence. Again, a value of $1.0$ denotes that we are dealing with interval uncertainty and a value of $0.0$ means that we now the cdf precisely.
	

\section{Numerical Simulation}
\label{sec:Results}
Consider the following linear dynamic system driven by Gaussian white noise:
\begin{eqnarray} \label{ex1_model}
\dot{x} + a_1 x &=& \Gamma_1(t) \\
p(t_0, x_0) &=& \mathcal{N}(x_0;1.1,2.42)
\end{eqnarray}
where the autocorrelation function of the noise is $E[\Gamma_1(t)\Gamma_1(\tau)] = q_1^2 \delta(t-\tau)$, and both $a_1, q_1$ are epistemic variable, described by the following body of evidence:
\small
\begin{eqnarray}
a_1 &\sim& \bigg\{ \left( [0.86, 0.9], 0.2 \right) ~,~ \left([0.89, 0.96], 0.8 \right)\bigg\} \\
q_1 &\sim& \bigg\{ \left( [0.2, 0.3], 0.6 \right) ~,~ \left([0.3, 0.4], 0.4 \right) \bigg\}
\end{eqnarray}
\normalsize

We are interested in finding the quantity of interest $P(t = 2, Y \le -0.5)$ as well as construct the entire cdf, $P(t = 2, Y \le y)$, after $t = 2$ sec. 

Here, the exact conditional probability density function is Gaussian,
\begin{eqnarray}\label{exact_pdf}
p(t,x|a_1,q_1) = \mathcal{N}(x ; \mu(t), \sigma^2(t)|a_1,q_1)
\end{eqnarray}
because the model is linear and the initial condition and the process noise are normally distributed.

Thus, we are interested in finding the induced DS structures for the moments of the normal distribution in Eq.\eqref{exact_pdf}. The time evolution of moments through which the epistemic uncertainty is propagated is given by:
\begin{eqnarray}\label{ex1_momeq}
\dot{m}_1 &=& -a_1 m_1 \\
\dot{m}_2 &=& -2 a_1 m_2 + q_1^2 \nonumber \\
&& m_1(0) = 1.1 \quad \mathrm{and} \quad m_2(0) = 2.42 \nonumber
\end{eqnarray}

\begin{figure*}
\centering
\subfigure[DS-structure for $\{\mu,\sigma^2\}$]{\includegraphics[width=2.8in]{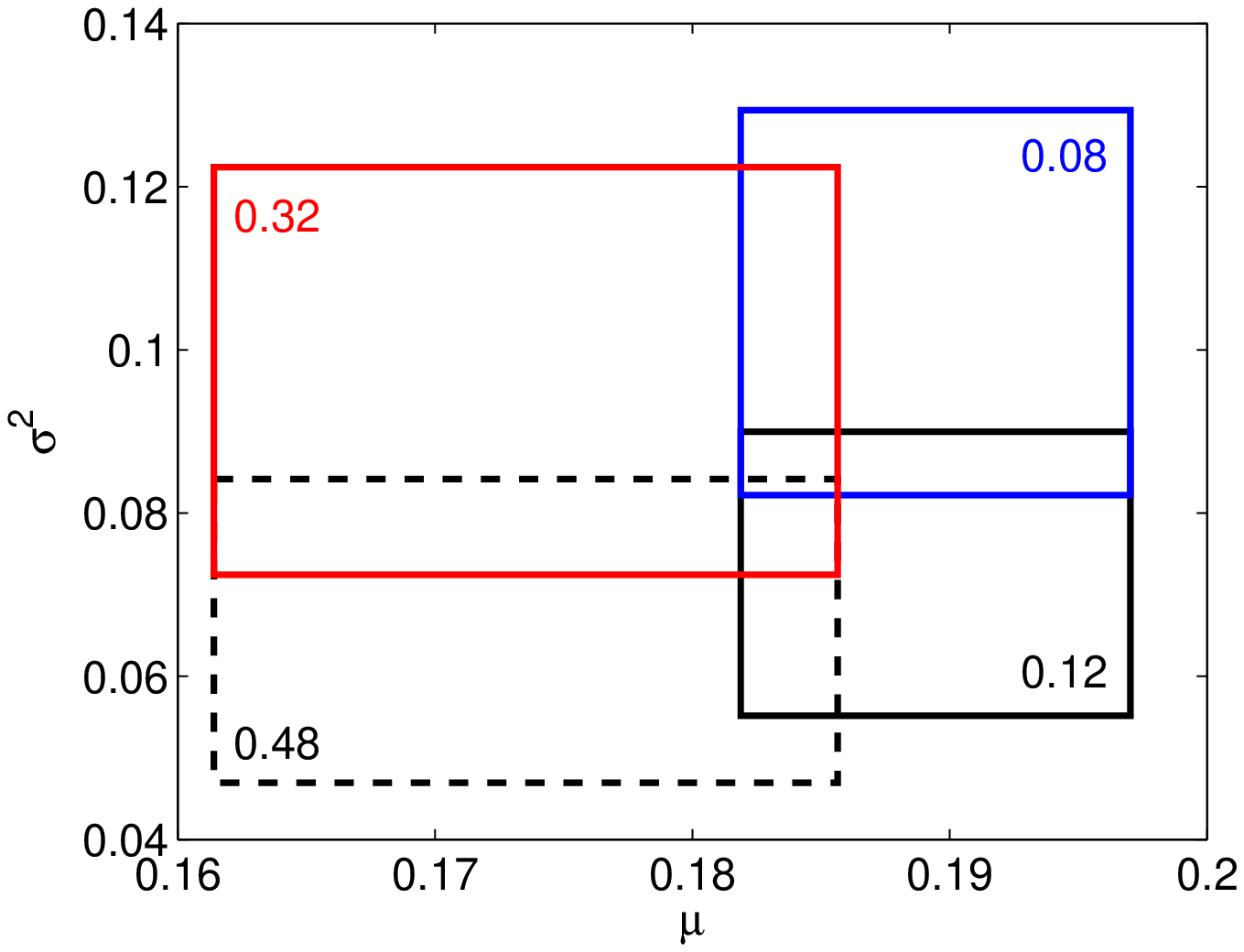}} 
\subfigure[DS-structure for $P(t = 2, Y \le y | a_1, q_1)$]{\includegraphics[width=2.8in]{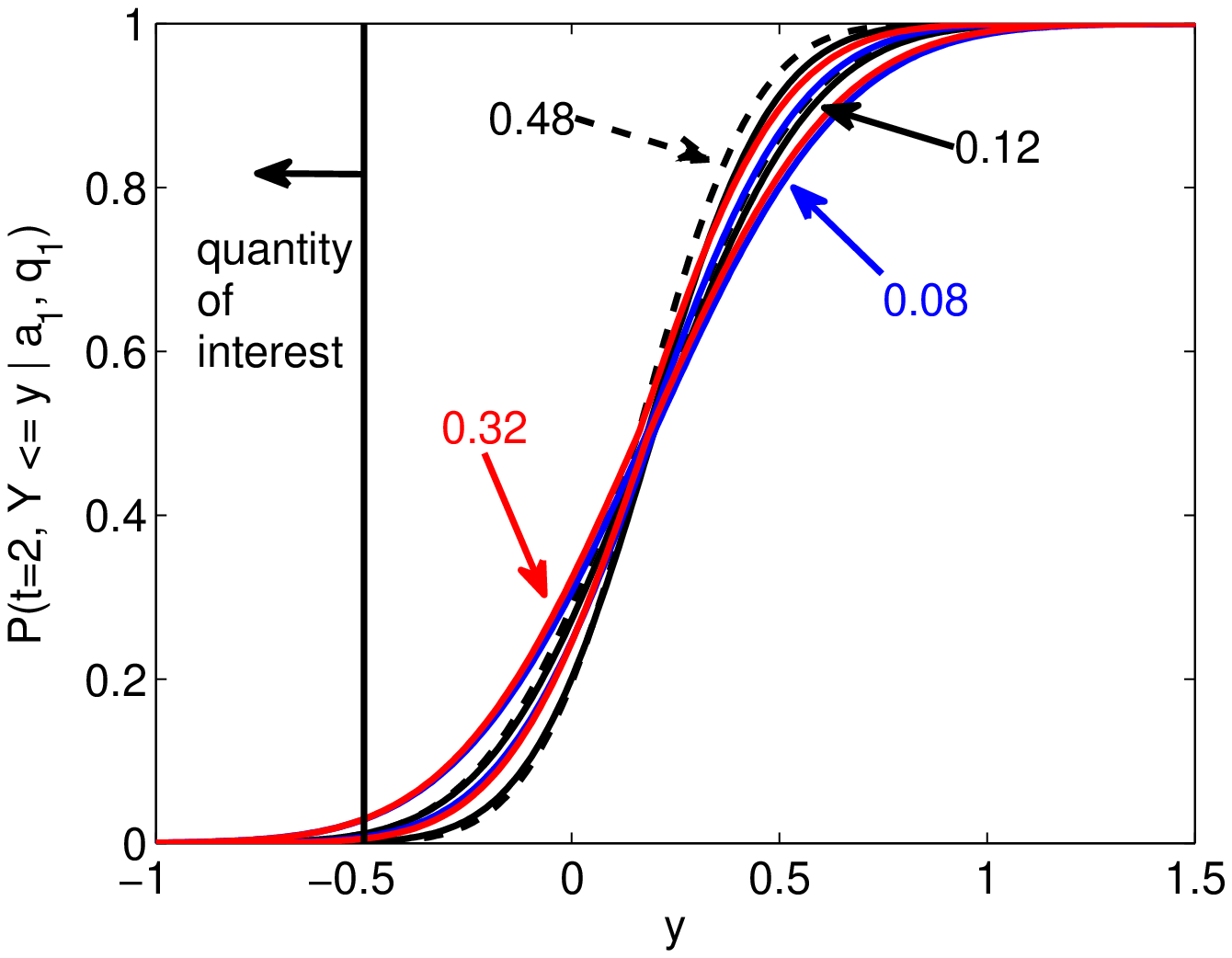}} 
\subfigure[DS-structure for $P(t=2, Y \le -0.5| a_1, q_1)$]{\includegraphics[width=2.8in]{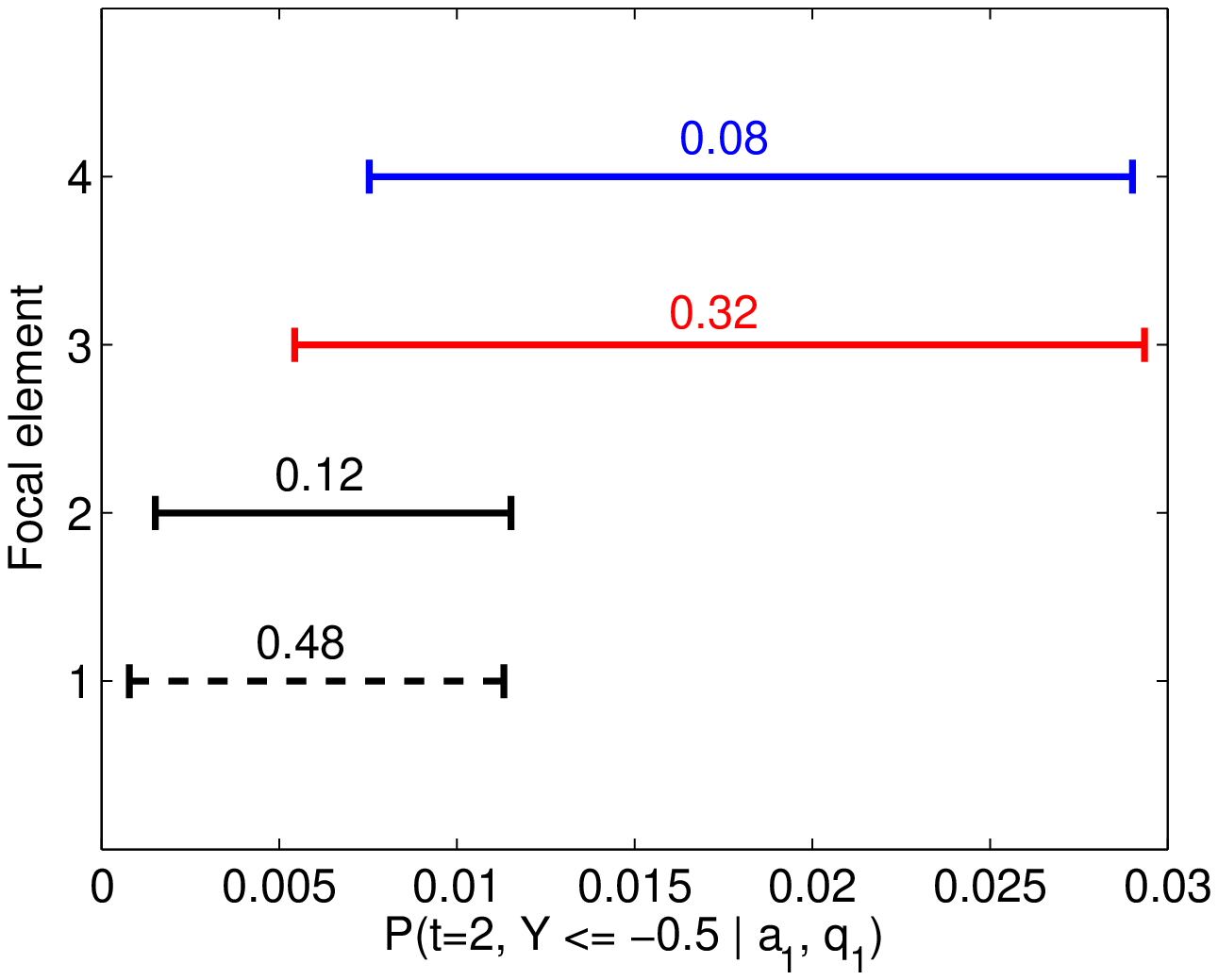}} 
\subfigure[P-box and pignistic cdf for $P(t= 2, Y \le -0.5| a_1, q_1)$]{\includegraphics[width=2.8in]{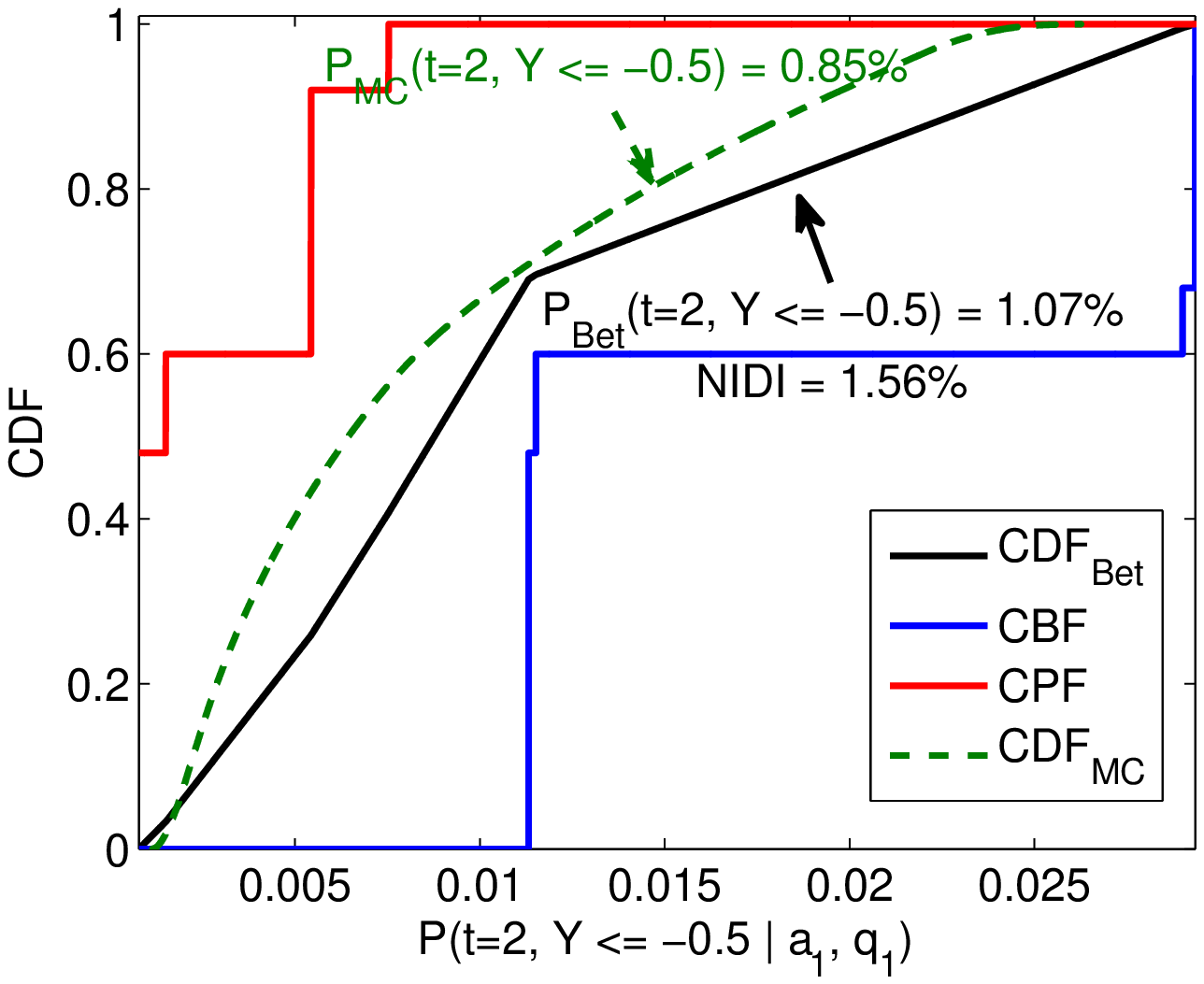}} 
\subfigure[Histogram for $P(t=2, Y \le -0.5| a_1^*, q_1^*)$]{\includegraphics[width=2.8in]{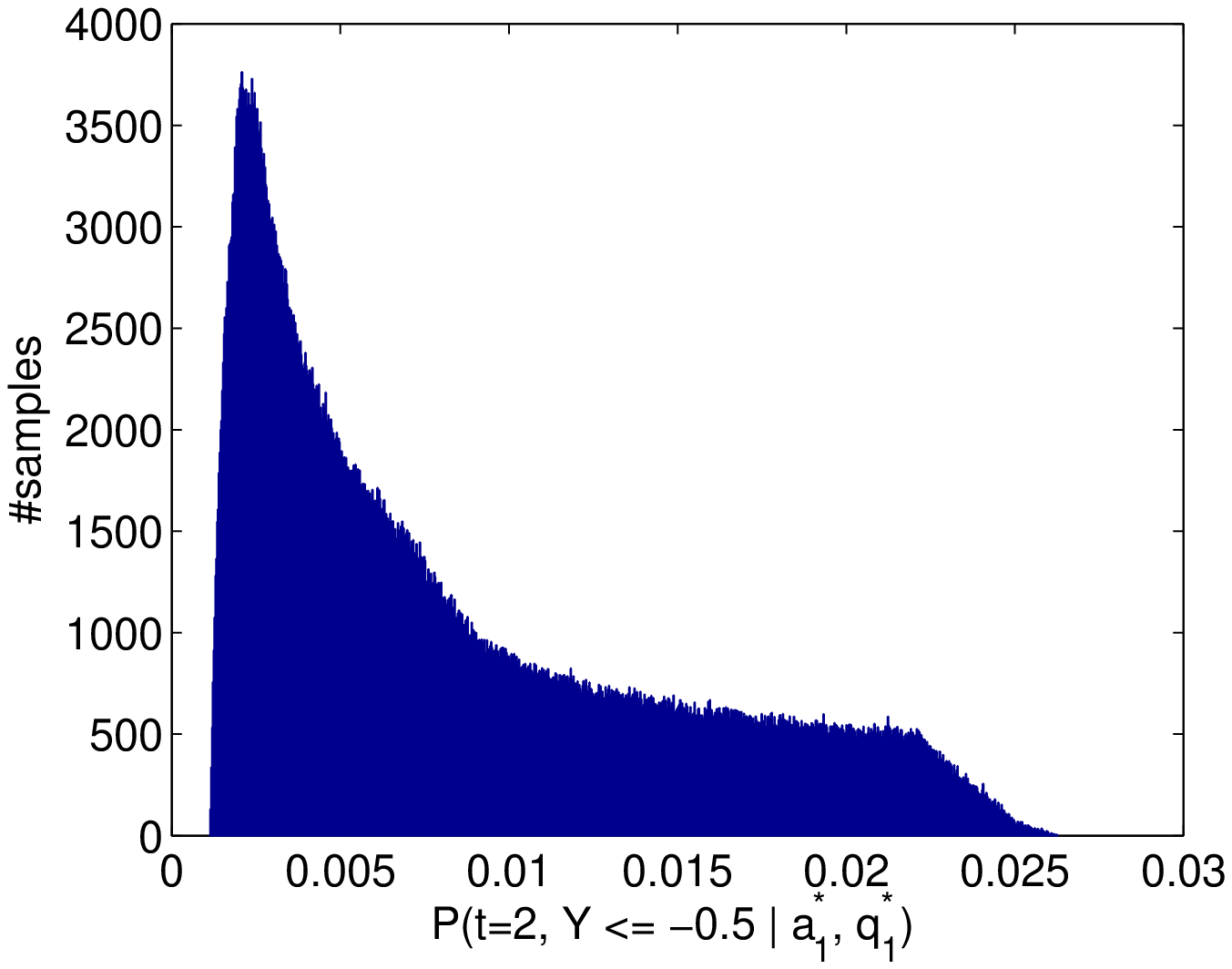}} 
\subfigure[Constructed CDF for $y$ and Ignorance Function]{\includegraphics[width=2.8in]{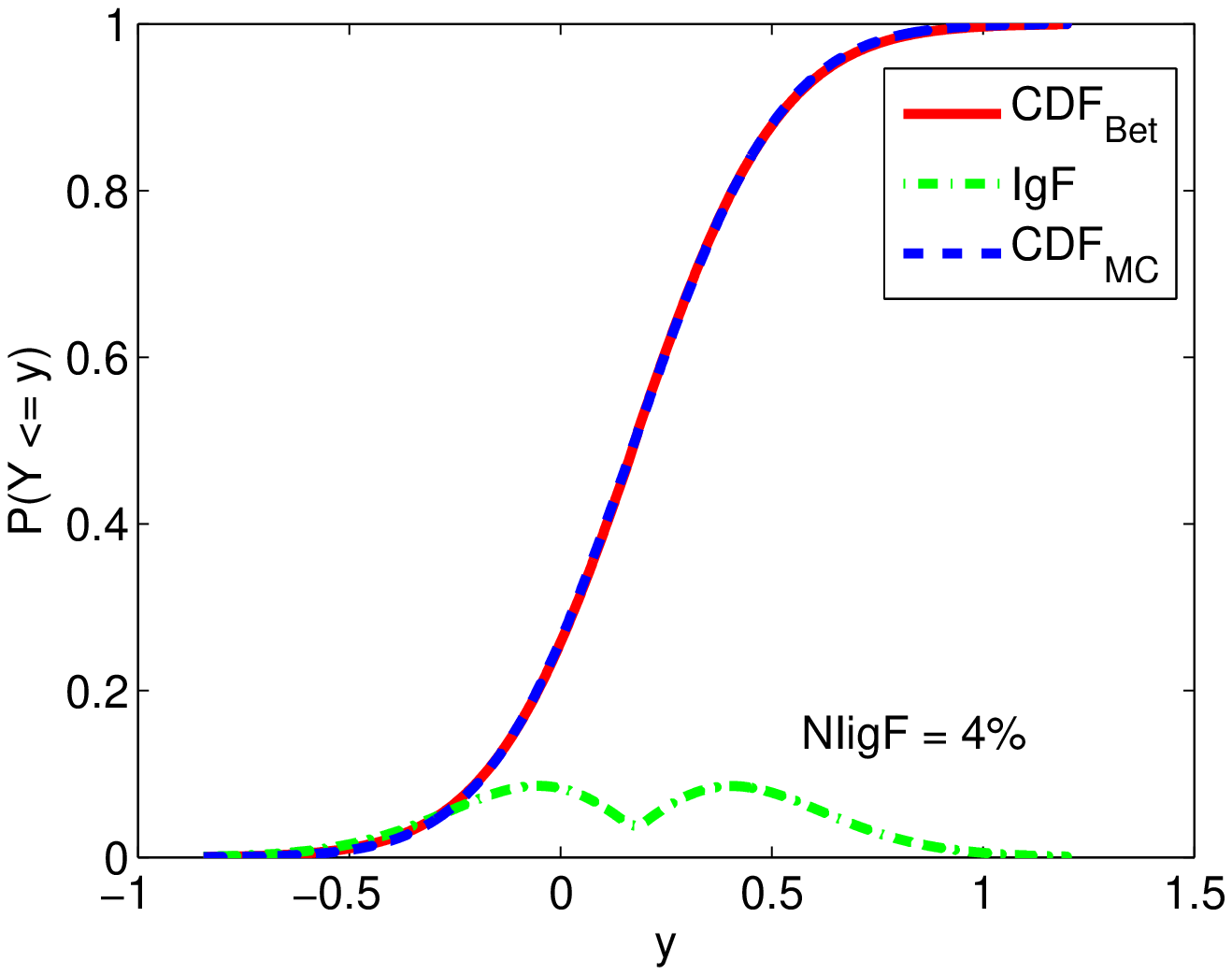}} 
\caption{Numerical simulation: DS structures, pignistic cdfs and ignorance function after $t = 2$ sec}\label{fig:DSex1}
\end{figure*}

Using the PCE - Bernstein form described in Section \ref{sec:PCEbernstein} to propagate interval uncertainties through the ODE in Eq.\eqref{ex1_momeq}, and solving for the central moments we find the DS structure for $\mu$ and $\sigma^2$ after $t = 2$ sec, shown here in Table \ref{tab:ex1_DS} and graphically represented in Fig.\ref{fig:DSex1}a.

\begin{table*}[tbp]
	\small
  \centering
  \caption{The induced DS structure for $\{\mu, \sigma^2\}$}
    \begin{tabular}{rrr}
    \addlinespace
    \toprule
          & $m_{a_1}([0.86,0.9])=0.2$ & $m_{a_1}([0.89, 0.96])=0.8$ \\
    \midrule
    $m_{q_1}([0.2,0.3])=0.6$ & $m_{\{\mu,\sigma^2\}}\bigg({[0.182, 0.197],[0.055, 0.090]}\bigg)=0.12$ & $m_{\{\mu,\sigma^2\}}\bigg({[0.161, 0.186],[0.047, 0.084]}\bigg)=0.48$ \\
    $m_{q_1}([0.3,0.4])=0.4$ & $m_{\{\mu,\sigma^2\}}\bigg({[0.182, 0.197],[0.082, 0.129]}\bigg)=0.08$ & $m_{\{\mu,\sigma^2\}}\bigg({[0.161, 0.186],[0.076,0.122]}\bigg)=0.32$ \\
    \bottomrule
    \end{tabular}
  \label{tab:ex1_DS}
\end{table*}

The DS structure over p-boxes is obtained by using the envelope property in Eq.\eqref{envelope_prop}. The focal elements and their bpa for $P(t = 2, Y \le y | a_1, q_1)$ are presented in Fig.\ref{fig:DSex1}b. From this structure it is easy to obtain the DS structure for $P(t = 2, Y \le -0.5 | a_1, q_1)$ as it is shown in Fig.\ref{fig:DSex1}c. Now, using the pignistic pdf for $P(t = 2, Y \le -0.5 | a_1, q_1)$ one can construct an estimate for the quantity of interest using Eq.\eqref{pigCdf} as well as provide a measure of confidence for this estimate:
\small
\begin{eqnarray}
P_{Bet}(t=2, Y \le -0.5) = 1.07\% \quad \mathrm{with} \quad NIDI = 1.56\%
\end{eqnarray}
\normalsize

Eq.\eqref{pigCdf} can be used to construct a singleton cdf for $y$ to take an action using the expected utility theory. Also using Eq.\eqref{igorance_fun} we can obtain an ignorance function to indicate the point-wise confidence in the constructed cdf as well as provide a scalar measure of trust, Eq.\eqref{ig_meas}, in using this cdf in decision making. Both the constructed cdf and the ignorance function are presented in Fig.\ref{fig:DSex1}f.

Just for comparison purposes the probability theory has be used to obtain the total probability of $y$ under the assumption that the epistemic variables vary randomly. Using Laplace's Principle of Insufficient Reasoning and we can transform the epistemic variables $a_1$ and $q_1$ into the aleatory variables $a_1^*$ and $q_1^*$. For this, we use the pignistic transformation to find the following pignistic pdfs:
\begin{eqnarray}\label{param_pdf}
p_{Bet}(a_1^*) &=& 0.2 \mathcal{U}(0.86, 0.9) + 0.8 \mathcal{U}(0.89 0.96) \\
p_{Bet}(q_1^*) &=& 0.6 \mathcal{U}(0.2, 0.3) + 0.4 \mathcal{U}(0.3, 0.4) \nonumber
\end{eqnarray}

With this transformation the uncertainty in the model in Eq.\eqref{ex1_model} is represented solely by probabilities. Thus an estimate for the quantity of interest is calculated as in Eq.\ref{prob_MC}. 

\begin{strip}
\rule[0in]{3.3in}{.001in}
\small
\begin{eqnarray}\label{prob_MC}
P_{MC}(t=2, Y \le -0.5) &=& \int_{V} P(t = 2, Y \le -0.5 | a_1^*, q_1^*) p_{Bet}(a_1^*) p_{Bet}(q_1^*) \mathrm{d}a_1^*\mathrm{d}q_1^* \\
&=& \int_{V} \bigg[\int_{-\infty}^{-0.5} \mathcal{N}(x; \mu, \sigma^2 | a_1^*, q_1^*) \mathrm{d}x \bigg] p_{Bet}(a_1^*) p_{Bet}(q_1^*) \mathrm{d}a_1^*\mathrm{d}q_1^* \nonumber \\
&=& 0.85\% \nonumber
\end{eqnarray} 
\normalsize
\begin{flushright}
\rule[0in]{3.3in}{.001in}
\end{flushright}
\end{strip}

In Fig.\ref{fig:DSex1}e it is plotted the histogram for $P(Y \le -0.5 | a_1^*, q_1^*)$ given $1$ million samples drawn from the pdfs in Eq.\eqref{param_pdf}. Also the empirical cdf for $P(Y \le -0.5 | a_1^*, q_1^*)$ is shown in Fig.\ref{fig:DSex1}d and the total cumulative probability for $y$ is presented in Fig.\ref{fig:DSex1}f.

The probabilistic estimate for the quantity of interest differs from the one given by the present approach due to the assumption that the model parameters vary randomly. Since we have used the pignistic transformation to obtain a pdf for the model parameters and then propagate this pdf through the dynamic system, the concentration in the probability mass seen in Fig.\ref{fig:DSex1}e cannot be explained given the evidence we have started with. This argument is nicely explained for a simple example by Ferson in Ref.\cite{Ferson1996}. More over the $NIDI$ measure indicates how much can we trust our estimate given our lack of knowledge about the model parameters. The threshold that indicates if the estimate can be trusted can only depend on the magnitude of the consequences is the state of the system drops below $-0.5$.

The two cdf for $y$ in Fig.\ref{fig:DSex1}f are slightly different, however for the cdf constructed using the present method we also include an ignorance function as a point-wise measure of confidence and a number, $NIigF$, which similar with $NIDI$. Here, while the probabilistic approach offers no other alternative just to use the expected utility theory to take an action, our approach through the use of the ignorance function and the $NIigF$ measure can postpone the decision making if the amount of ignorance is high. The use of ignorance function and $NIigF$ remains to be studied in the decision making problem, and is not the purpose of the current paper.

\section{Conclusions}
\label{sec:Conclusions}
A new framework has been presented to propagate both aleatory and epistemic uncertainty through dynamic systems with stochastic forcing. This has been achieved by using a second-order uncertainty model to propagate both types of uncertainty. 

First an approximation to the probability density function of the response is assumed and it is used in the moment closure scheme to find the corresponding time evolution of the moments. The epistemic uncertainty is mapped through these equations and the final response of the system is modeled as a Dempster Shafer structure on probability boxes.

The paper incorporates a previous work of the author to propagate interval uncertainties through ordinary differential equations using the polynomial chaos expansion and the Bernstein form. However, other methods such as Taylor models and interval arithmetic can be used to propagated the focal elements through the moment equations.

It is shown that the Demster Shafer structures obtained can be used in constructing estimates of quantity of interests, such as probabilities of failure, and also in constructing entire cumulative density functions which can be used in decision making. In addition, an ignorance function and a scalar measure can be computed in order to hint the confidence that one can have in the estimates given the initial body of evidence concerning the model parameters. For now one may choose not to pursue with the decision making if the scalar measure is over a certain threshold. The use of the ignorance function in conjunction with the expected utility theory is set as future work.
\\
\\
\textbf{Acknowledgment:} 
\textit{This work was supported under Contract No. HM1582-08-1-0012 from ONR.}

\small
\bibliographystyle{plain}
\bibliography{Fusion2010}

\end{document}